# Inhomogeneity and complexity measures for spatial patterns

R. Piasecki[a], M.T. Martin[b], A. Plastino[b,c,*]

[a]*Institute of Chemistry, University of Opole, Oleska 48, PL 45052 Opole, Poland*
[b]*Argentina National Research Council (CONICET), Argentina*
[c]*Physics Department, National University La Plata, C.C. 727, (1900) La Plata, Argentina*

**Abstract**

In this work, we examine two different measures for inhomogeneity and complexity that are derived from non-extensive considerations à la Tsallis. Their performance is then tested on theoretically generated patterns. All measures are found to exhibit a most sensitive behaviour for Sierpinski carpets. The procedures here introduced provide us with new, powerful Tsallis' tools for analysing the inhomogeneity and complexity of spatial patterns.

*PACS:* 05.90.+m

*Keywords:* Non-extensive statistics; Complexity; Spatial patterns

## 1. Introduction

In spite of its great success, the statistical mechanics paradigm based on the Boltzmann–Gibbs entropic measure seems to be inadequate to deal with many interesting physical scenarios [1–3]. Astronomical self-gravitating systems constitute an important illustrative example of these difficulties [4].

In this paper, we are concerned with measures of spatial inhomogeneity and complexity for different length scales. A recent study [5] has advanced a quantitative characterisation of morphological features of a material system. This characterisation is based upon a normalised information (entropic) measure, more general than both (i) the so-called local porosity entropy [6] and (ii) the "configuration entropy" [7], two concepts that have been shown to be connected in Ref. [8]. Here we wish to introduce Tsallis' generalised measures into this area of endeavour. We will show that they provide us with a quite useful research tool that allows for fruitful insights concerning both inhomogeneity and complexity of spatial patterns.

---

*Corresponding author. Physics Department, National University La Plata, C.C. 727, (1900) La Plata, Argentina. Fax: +54-2258238.
*E-mail address:* plastino@venus.fisica.unlp.edu.ar (A. Plastino).

## 2. Entropic measures

### 2.1. Recapitulation of the microcanonical formalism and of the averaging procedure

We deal here mainly with the two entropic measures (i) $S_\Delta$ and (ii) $S_\Delta(PO)$. The first one (see formula (3a)) refers to indistinguishable finite-sized objects (FSOs) and is fully described in Ref. [9]. The second measure (see formula (3b)) refers to the simpler case of "point" objects (POs) and was used in Refs. [10,11] under a different guise (it was called "$f(S)$" in these references). The method of Ref. [9] (with regard to $S_\Delta$) is exact when applied to the evaluation of the spatial inhomogeneity of identical FSOs (at different length scales). The approach of Refs. [10,11] can be regarded as exact for investigating the degree of inhomogeneity in a phase space. When the entropic measures are used for the same binary image, that is, (1)-pixel = finite-sized object, exact measure, and (2)-pixel $\cong$ point marker as a punctual object, approximated measure that is, however, qualitatively correct, a simple relation between them can be established (see Eq. (3c)). This formula clearly shows that an FSO-measure "sees" its surroundings (regarded as a second phase) as composed of finite-sized objects. Thus, this measure can be applied with reference to a two-phase system of FSOs. The PO-measure, instead, is better suited for the description of punctual inclusions in a continuous medium. For more details see below (for FSOs) and also Appendix A (for POs).

Consider a mixture of non-interacting and of equal size ($1 \times 1$ in pixels) black and white objects (it should be noted that, arguably, these "particles" interact with each other through mutual exclusion). We talk of, respectively, black and white pixels. For a given $L \times L$ grid, let us have $n$ black pixels ($0 < n < L^2$) pixels and $m = L^2 - n$ white pixels, distributed in square, non-overlapping and numerable $\chi = (L'/k)^2$ lattice cells of size $k \times k$ (no pores). We will use the abbreviation $L' \equiv lL$, where $l$ is a natural number employed to form, by periodic repetition of an initial arrangement, a final pattern of size $lL \times lL$ (see the explanation given below, after formula (3a)). For each length scale $k$ we assume standard constraints for the cell occupation numbers, i.e., $n_1 + n_2 + \cdots + n_{\chi(k)} = n$, and, correspondingly, $m_1 + m_2 + \cdots + m_{\chi(k)} = m$ with $m_i(k) + n_i(k) = k^2$ for $i = 1, 2, \ldots, \chi(k)$.

Consider all distributions of these objects with fixed occupation numbers as a kind of scale-dependent configurational *macrostate*, described by the set $\{n_i(k)\} \equiv \{n_1, n_2, \ldots, n_{\chi(k)}\}$ (for black pixels). A corresponding macrostate for white pixels $\{k^2 - n_1, k^2 - n_2, \ldots, k^2 - n_{\chi(k)}\}$ is automatically at hand. We can limit ourselves to a black pixels' language. For a given length scale $k$, any macrostate can be realized by a number of distinguishable arrangements of $n$ black pixels associated with the $1 \times 1$ lattice cells, i.e., a kind of "equally likely" configurational *microstates*, with $n_i = 0$ or 1. Setting the Boltzmann constant $k_B = 1$ we use the standard definition of configurational (Boltzmann microcanonical) entropy

$$S(k) = \ln \Omega(k) \equiv \ln \left[ \prod_{i=1}^{\chi} \binom{k^2}{n_i} \right], \tag{1}$$

where $\Omega(k)$ is the total number of distinguishable spatial arrangements of the objects in $\chi(k)$ cells of size $k \times k$, each containing $n_i(k)$ black pixels and $m_i(k)$ white pixels.

The number of distinguishable spatial arrangements of our objects increases when they are mixed. For every length scale $1 < k < L$, there is *a special set* of the "most spatially ordered" object-distributions containing $\Omega_{\max}(k)$ configurations. These are distinguished by the condition $|n_i(k) - n_j(k)| \leqslant 1$ (holding for each pair $i \neq j$). Every configuration belonging to such a *special set* represents a *reference configurational macrostate* $\{n_i(k)\}_{\text{RCM}}$ having the highest possible value of configurational entropy $S_{\max}(k)$. On the other hand, for $k = 1$ and for $k = L$ we have always $S_{\max}(1) = S(1) = 0$ and $S_{\max}(L) = S(L) \neq 0$. In general, for a given $\{n_i(k)\}_{\text{RCM}}$

$$S_{\max}(k) = \ln \Omega_{\max}(k) \equiv \ln \left[ \binom{k^2}{n_0}^{\chi - r_0} \binom{k^2}{n_0 + 1}^{r_0} \right] \tag{2}$$

with $r_0 = n \bmod \chi$, $r_0 \in \{0, 1, \ldots, \chi - 1\}$, and $n_0 = (n - r_0)/\chi$, $n_0 \in \{0, 1, \ldots, k^2 - 1\}$.

We shall concentrate our efforts on ascertaining the dependence of the spatial inhomogeneity of the mixture *on the length scale $k$* by using an information measure to be defined below. In order to evaluate, for each $k$, the deviation of the macrostate $\{n_i(k)\}$ represented by an actual configuration from the appropriate $\{n_i(k)\}_{\text{RCM}}$ one it is natural to consider the difference $S_{\max}(k) - S(k)$. Averaging this difference over the number of cells $\chi(k)$, we obtain a relative measure (per cell) of the spatial inhomogeneity of our mixture

$$S_\Delta(k) = \frac{S_{\max} - S}{\chi} \equiv \frac{r_0}{\chi} \ln\left(\frac{k^2 - n_0}{n_0 + 1}\right) + \frac{1}{\chi} \sum_{i=1}^{\chi} \ln\left[\frac{n_i!(k^2 - n_i)!}{n_0!(k^2 - n_0)!}\right]. \tag{3a}$$

This averaging procedure is necessary in order to guarantee the validity of a crucial property that an information measure must possess if we wish to be able to perform calculations at every length scale: if a final pattern of size $lL \times lL$ (where $l$ is a natural number) is formed by the periodic repetition of an initial arrangement of size $L \times L$, then the value of the measure at a given length scale $1 \leqslant k \leqslant L$ commensurate (i.e., when $k$ is an integer divisor of $L$) with the side length $L$ being kept *unchanged* under the replacement $L \leftrightarrow lL$, since it also causes $n \leftrightarrow l^2 n$, $\chi \leftrightarrow l^2 \chi$, $r_0 \leftrightarrow l^2 r_0$, keeping unchanged both $n_0$ and the corresponding $n_i$. To overcome the problem of an incommensurate length scale it is sufficient to determine a number $l$ such that $lL \bmod k = 0$ and replace then the initial arrangement of size $L \times L$ by the periodically created one of size $lL \times lL$. We define then $S_\Delta(k, L, n) \equiv S_\Delta(k, lL, l^2 n)$. According to this definition, every binary pattern can be treated as spatially homogeneous at length scales $k = 1$ and $L$.

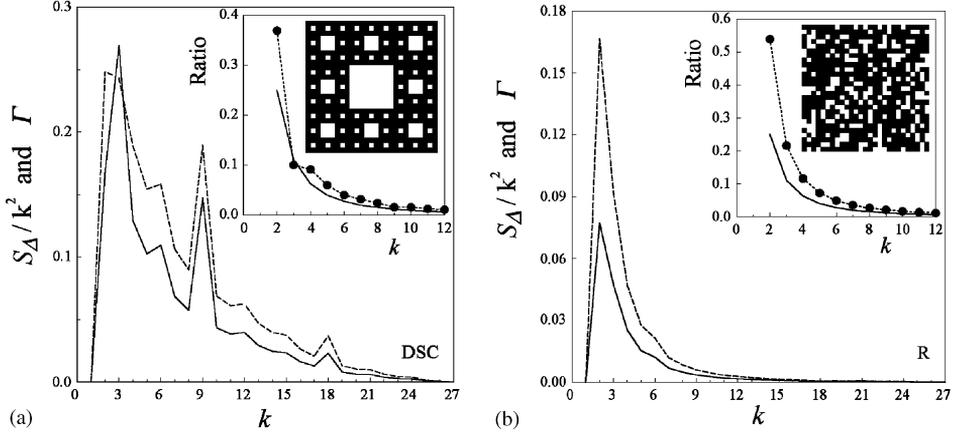

Fig. 1. Modified $S_\Delta(k)/k^2$-measure (solid line) and complexity measure $\Gamma(k)$ (dashed line), plotted as a function of the length scale for the binary patterns depicted in the insets. The correlation between the two measures is also shown in the insets: full circles correspond to the $\Gamma(k)/S_\Delta(k)$ ratio, while the solid line depicts the $1/k^2$ function. (a) DSC(3,8,3) pattern and (b) R pattern.

The corresponding formula for PO-measures can be obtained in a similar way (see (A1–A3) in Appendix A) and reads

$$S_\Delta(k; \text{PO}) = -\frac{r_0}{\chi} \ln(n_0 + 1) + \frac{1}{\chi} \sum_{i=1}^{\chi} \ln \left[ \frac{n_i!}{n_0!} \right] . \tag{3b}$$

The exact relation between the two measures (applied to the same binary pattern) is given by

$$S_\Delta(k) = S_\Delta(k; \text{PO}) + \frac{r_0}{\chi} \ln(k^2 - n_0) + \frac{1}{\chi} \sum_{i=1}^{\chi} \ln \left[ \frac{(k^2 - n_i)!}{(k^2 - n_0)!} \right] . \tag{3c}$$

In recent studies [12,13] a measure of "complexity" was considered that can be expressed in terms of order/disorder. Adapted to our spatial case, the simplest form of this measure can be written as

$$\Gamma(k) = \begin{cases} 0 & \text{for } k = 1, L, \\ \Delta(1 - \Delta) & \text{for } 1 < k < L, \end{cases} \tag{4}$$

where $\Delta \equiv S(k)/S_{\max}(k)$. Note that this "(length-scale)-dependent" form differs from the original one given in Ref. [13] (see also Ref. [14]) in the kind of entropies ($k$-dependence) that enter the entropic ratio. The two binary patterns depicted in the insets of Fig. 1a and b are used for testing purposes. For a deterministic Sierpinski carpet [9] the notation DSC$(a,b,c)$ refers to an initial square lattice of $L \times L$ cells, with $L = a^c$, divided into $a^2$ sub-squares, with only $b$ conserved according to a deterministic rule. Such a decimation procedure is repeated on each conserved sub-square, and so on, $c$ times. The second pattern refers to a low structured random arrangement (called an R pattern) of the same numbers $n = 512$ (black) and $m = 217$ (white) pixels, as, for instance, DSC(3,8,3).

There is a clear correlation between $S_\Delta(k)/k^2$, on the one hand, and $\Gamma(k)$ on the other one (see Fig. 1a and b). This behaviour can be easily understood on the basis of the rough estimation for the ratio $S_\Delta(k)/k^2$ to $\Gamma(k)$ given in Appendix B.

## 2.2. Extension of the formalism to a generalised setting and examples

Focus attention now upon Tsallis generalised information measure [1]. It is easy to see that

$$S_q = \frac{1}{q-1}\left[1 - \sum_{i=1}^{W} p_i^q\right] \xrightarrow[\text{for each } i \; \lim p_i = 1/W]{} \frac{W^{1-q} - 1}{1-q} \xrightarrow[\lim q=1]{} \ln W, \qquad (5a)$$

or

$$S_q = \frac{1}{q-1}\left[1 - \sum_{i=1}^{W} p_i^q\right] \xrightarrow[\lim q=1]{} -\sum_{i=1}^{W} p_i \ln p_i \xrightarrow[\text{for each } i \; \lim p_i = 1/W]{} \ln W, \qquad (5b)$$

where $W$ is the total number of possible microstates ($W \equiv \Omega$ in our notation for configurational microstates), $\{p_i\}$ the associated probabilities, and $q$ the *non-extensivity* (real) parameter. For convenience, the constant Boltzmann's factor is set equal to unity.

At this point our investigation begins. We ask now *whether there exists a well-defined q-extension of the information measure given by Eq.* (3a). We require also that it should retain the above referred to crucial property, i.e., the one that allows for calculating the measure at every length scale. What motivates this type of generalisation is the expectation that the non-linearity of the $q$-measure will reveal new features regarding the comparison of inhomogeneity and complexity measures for POs and FSOs. Note also that we can use (in the reverse direction) each of the middle expressions of (5a) and (5b).

### 2.2.1. A-approach

Let us denote as (**A**) the (5a)-approach. Considering the appropriate equally likely configurational microstates (with $n_i = 0$ or 1) for a given macrostate, for both (i) $p_i(k) = 1/\Omega(k)$ and (ii) the "reference macrostate" $p_{i,\max}(k) = 1/\Omega_{\max}(k)$, we obtain in natural fashion

$$S_q(k; \mathbf{A}) = \frac{\Omega(k)^{1-q} - 1}{1-q}, \qquad (6a)$$

$$S_{q,\max}(k; \mathbf{A}) = \frac{\Omega_{\max}(k)^{1-q} - 1}{1-q} \qquad (6b)$$

and

$$S_{\Delta,q}(k; \mathbf{A}) = \frac{1}{\chi} \frac{\Omega_{\max}(k)^{1-q} - \Omega(k)^{1-q}}{1-q}. \qquad (7)$$

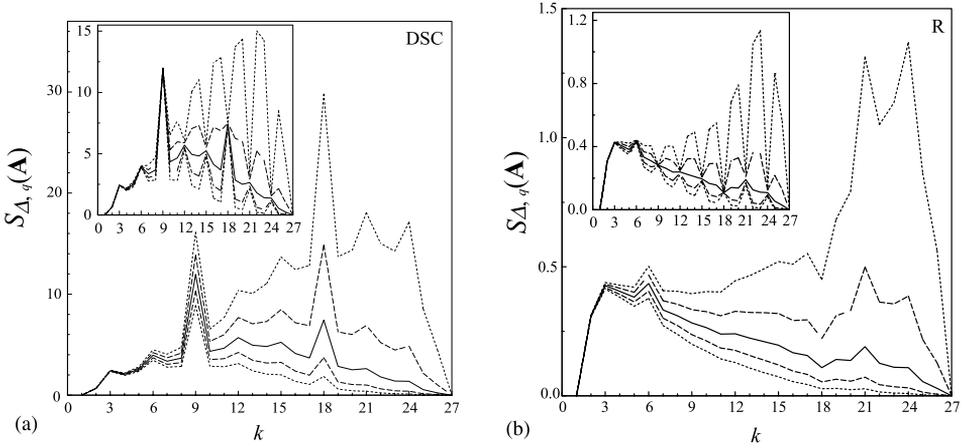

Fig. 2. Entropic $q$-measure (Cf. Eq. (7), **A** approach) for FSOs, as a function of the length scale, for a constant number of cells (in the insets, for an optimal number of cells). Short-dashed lines (bottom for $q$ and upper for $1/q$) correspond to $q = 1.00001$, while long-dashed lines (bottom ($q$) and upper ($1/q$)) are drawn for $q = 1.000005$. We depict also the corresponding Boltzmann's measure (Cf. Eq. (3a)) for $q = 1$ (solid line). (a) DSC(3,8,3) and (b) R pattern.

A word of caution is needed here. For a fixed configurational macrostate $\{n_i(k=1)\}$ there is only one proper microstate and $\Omega(1) = \Omega_{\max}(1) = 1$. In turn, for a given $\{n_i(k=L)\}$, we have $\Omega(L) = \Omega_{\max}(L) \neq 0$. Thus, in both limit cases we have $S_{\Delta,q}(k=1$ or $L;\mathbf{A}) \equiv 0$, as expected. However, for any configurational macrostate $\{n_i(1 < k < L)\}$, Tsallis' measure depends on the number of cells $\chi(k)$ in such a way that results for (i) a constant cell number (that is, when $L' = kL$ and $\chi \equiv L^2 =$ constant) *will differ* from those for (ii) the optimal one (that is, when $L' = lL$ and $\chi \neq$ constant). This is not what happens with the (correct) behaviour of the measure $S_\Delta(k)$ given by Eq. (3a). Such a situation is illustrated in Fig. 2. In both cases the curves corresponding to $q > 1$ ($q < 1$) converge to the case $q = 1$ from opposite sides.

An additional interesting fact is to be pointed out here: $S_{\Delta,q}$ given by Eq. (7) *exhibits the same features* (upper curves for $q < 1$ and bottom ones for $q > 1$) displayed, with reference to the case $q = 1$ (see Fig. 1 of Ref. [15]), by the Kolmogorov–Sinai–Tsallis entropy (evaluated numerically). The 1D window of length $N$ referred as *time* in Ref. [15] corresponds to our length scale $k$.

### 2.2.2. B-approach

Let us denote as (**B**) the (second) approach that employs Eq. (5b). Introducing *local* black pixels fraction $\gamma_i(k) = n_i/k^2$ for the $i$th cell of size $k \times k$, and the corresponding fraction for white pixels (namely, $1 - \gamma_i(k)$), the Shannon-like forms $S(k)$ and $S_{\max}(k)$ can be recast, using the Stirling approximation $\ln n! \sim n \ln n - n$, in the fashion

$$S(k;\mathbf{B}) \cong -k^2 \sum_{i=1}^{\chi} [\gamma_i \ln \gamma_i + (1-\gamma_i) \ln(1-\gamma_i)] \equiv k^2 \sum_{i=1}^{\chi} S_i(\gamma), \tag{8a}$$

$$S_{\max}(k; \mathbf{B}) \cong -k^2(\chi - r_0) \left[\varphi_0 \ln \varphi_0 + (1 - \varphi_0) \ln(1 - \varphi_0)\right]$$
$$- k^2 r_0 \left[\varphi_1 \ln \varphi_1 + (1 - \varphi_1) \ln(1 - \varphi_1)\right]$$
$$\equiv k^2 (\chi - r_0) S_0 + k^2 r_0 S_1 \tag{8b}$$

and finally,

$$S_{\Delta}(k; \mathbf{B}) = \frac{k^2}{\chi} \left[(\chi - r_0) S_0 + r_0 S_1 - \sum_{i=1}^{\chi} S_i(\gamma)\right], \tag{9}$$

where $\varphi_0 = n_0/k^2$ and $\varphi_1 = (n_0 + 1)/k^2$. We immediately recognise as familiar the forms $S_i(\gamma)$, $S_0$, and $S_1$. They are "local" entropies. The very small differences between these Shannon-like measures and the Boltzmann ones given by Eq. (3a) appear only at the smaller length scales. The corresponding curves are practically indistinguishable at our drawings' scale.

We can now generalise as above and obtain $S_q(k; \mathbf{B})$ by recourse to the Jackson's $q$-derivative [16] $D_q f(x) \equiv [f(qx) - f(x)]/[qx - x]$ (see, for instance, the work of Abe [17]). This derivative ascertains just how the function $f(x)$ "reacts" under dilatation of the $x$-coordinate, opening thus a wide door for getting thermodynamical insights [18]. Within the framework of such an approach the normalisation identity $\sum_{i=1}^{W} p_i \equiv 1$ is not really needed. This fact might have some possible bearings on the so-called incomplete normalisation for microcanonical ensembles [19].

We can apply the Jackson derivative to each pair of local concentrations, as for instance

$$-D_q \left[\gamma_i^x + (1 - \gamma_i)^x\right]\big|_{x=1} = \frac{1 - [\gamma_i^q + (1 - \gamma_i)^q]}{q - 1}, \tag{10}$$

thus getting

$$S_q(k; \mathbf{B}) = k^2 \sum_{i=1}^{\chi} \left\{\frac{1 - [\gamma_i^q + (1 - \gamma_i)^q]}{q - 1}\right\} \equiv k^2 \sum_{i=1}^{\chi} S_{i,q}(\gamma). \tag{11a}$$

Note that the appropriate expression for the configurational entropy of a mixture of black and white pixels, for a given length scale, is now obtained in terms of (i) the total number of cells and (ii) the local pixel fractions, *not in terms of W and $p_i$*. In a similar vein we find

$$S_{q,\max}(k; \mathbf{B}) = k^2 \left\{(\chi - r_0) \frac{1 - [\varphi_0^q + (1 - \varphi_0)^q]}{q - 1} + r_0 \frac{1 - [\varphi_1^q + (1 - \varphi_1)^q]}{q - 1}\right\}$$
$$\equiv k^2 \left[(\chi - r_0) S_{0,q} + r_0 S_{1,q}\right] \tag{11b}$$

and finally,

$$S_{\Delta,q}(k; \mathbf{B}) = \frac{k^2}{\chi} \left[(\chi - r_0) S_{0,q} + r_0 S_{1,q} - \sum_{i=1}^{\chi} S_{i,q}(\gamma)\right]. \tag{12}$$

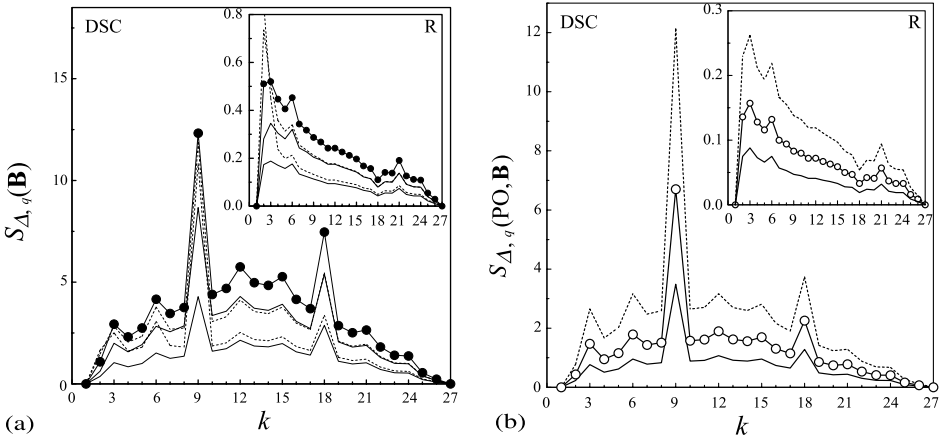

Fig. 3. Entropic $q$-measure in the **B** approach, as a function of the length scale for the DSC pattern (in the insets, for the R pattern). We depict also the corresponding Shannon's measure for $q = 1$: full circles, according to Eq. (9), and open ones according to Eq. (B3). (a) for FSOs (Cf. Eq. (12)), with $q = 2.5$ and 5, from top to bottom, respectively (solid lines). The associated $1/q$ counterparts are represented by dashed lines. (b) for POs (Cf. Eq. (13)) with a constant number of cells and $q = 1.1$ (solid line), and $1/q$ counterpart (dashed line).

For $q$ close to 1 the correctness of the above formulae was confirmed by using a Taylor expansion up to order four in the difference (Jackson) operator [20]. Fig. 3a depicts the typical behaviour of $S_{\Delta,q}(k;\mathbf{B})$ versus length scale $k$ for (i) two different $q$-parameters and (ii) for their associated inverse values $1/q$. Both the DSC(3,8,3) and the R patterns are here considered. The Tsallis form $S_{\Delta,q}(k;\mathbf{B})$ reduces to the corresponding Shannon one $S_\Delta(k;\mathbf{B})$ for $q \to 1$. For the sake of completeness we also give here the final formula for the PO-measure (Appendix A). In the case of the **B** approach it reads

$$S_{\Delta,q}(k;\text{PO},\mathbf{B}) = \frac{n}{\chi(q-1)} \left[ -(\chi - r_0)(n_0/n)^q - r_0((n_0+1)/n)^q + \sum_{i=1}^{\chi} (n_i/n)^q \right]. \tag{13}$$

Measure (13) exhibits a similar dependence on the number of cells $\chi(k)$, (either constant or optimal) as the measure given by Eq. (7). For the latter one this seems quite natural, since the measure $S_{\Delta,q}(k;\mathbf{A})$ a priori "neglects" any correlations between the $p_i$. On the other hand, the measure $S_{\Delta,q}(k;\text{PO},\mathbf{B})$ does not take into account the additional, length-scale dependent constraints for FSOs (like $n_i \leqslant k^2$), in contrast to what happens with the restricted local fractions $\gamma_i(k)$ and $1 - \gamma_i(k)$ incorporated into the measure $S_{\Delta,q}(k;\mathbf{B})$. Mathematically, these differences manifest themselves (i) in a global normalisation of the POs fractions, i.e., $\sum_{i=1}^{\chi} (n_i/n) \equiv 1$, and (ii) in a local one for FSOs fractions, $\gamma_i(k) + 1 - \gamma_i(k) \equiv 1$. Note that only in the latter case, considered globally, we have $\sum_{i=1}^{\chi} [\gamma_i + (1 - \gamma_i)] \equiv \chi$, as required in order that the "external" averaging (by the total number of cells $\chi(k)$) will remain effective.

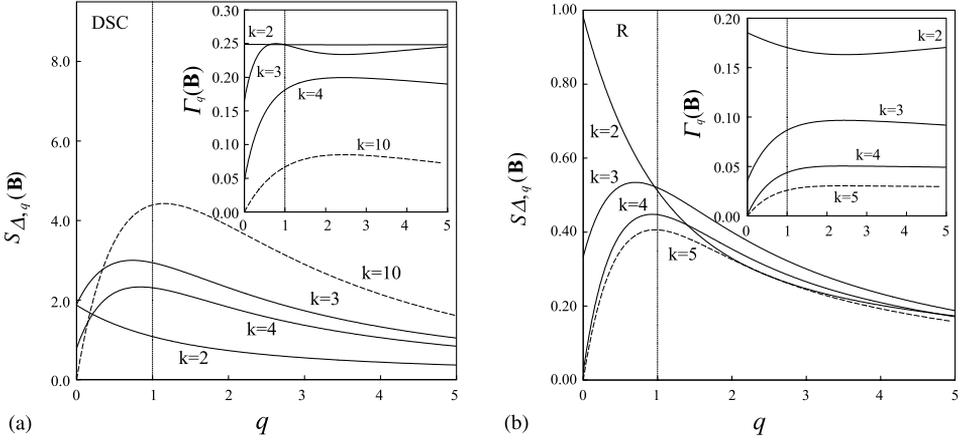

Fig. 4. Comparison of the $q$-measures $S_{\Delta,q}(\mathbf{B})$ and $\Gamma_q(\mathbf{B})$ (see inset), at fixed length scales, for the DSC pattern (a) and for the R-pattern (b). The dashed curves correspond to the smallest (first) length scale $k$ at which both measures vanish for $q \to 0^+$. Such behaviour indicates a lack of "empty" (with $n_i = 0$) cells of size $k \times k$. Moreover, for this length scale, and for larger ones as well, the duality $q \Leftrightarrow 1/q$ is clearly detected in Ref. [23]. Note that the corresponding Shannon's values of the measures appear at the crossing between our curves and the vertical short dashed line for $q = 1$.

The above feature of $S_{\Delta,q}(k; \mathrm{PO}, \mathbf{B})$ forces us to use it just for a constant number of cells. Fig. 3b depicts the typical behaviour of $S_{\Delta,q}(k; \mathrm{PO}, \mathbf{B})$, versus the length scale $k$, for a given $q$-parameter and also for its associated $1/q$-value, for both the DSC $(3,8,3)$ and the R patterns. Note that the behaviour displayed in Fig. 3b is quite different from that for FSOs in Fig. 3a. Also, the measure used in Fig. 3b is not symmetrical with respect to black and white pixels, as is the case for FSO-measures. We point out that the patterns investigated here are biased ones from the viewpoint of PO-measures: we have black pixel (not point) occupation numbers for a binary image, and $n_i$ never exceeds $k^2$. We should remember that all "true" point-objects can be placed into one (and the same) cell, independently of the cell's size.

### 2.2.3. Non-extensivity, inhomogeneity, and complexity measures

We are now in a position to make a comparison, for both FSOs (Fig. 4) and POs (Fig. 5), between (i) inhomogeneity and (ii) complexity measures, i.e., between $S_{\Delta,q}(k; \mathbf{B})$ and $S_{\Delta,q}(k; \mathrm{PO}, \mathbf{B})$, on the one hand, and the complexities $\Gamma_q(k; \mathbf{B})$ and $\Gamma_q(k; \mathrm{PO}, \mathbf{B})$, on the other one. The last two measures derive from Eq. (4), where we have replaced (in $\Delta$) the appropriate entropies with measures (11a), (11b) and (A.5,A.6), respectively.

The DSC pattern is the subject of Fig. 4a. We depict there the $q$-measure $S_{\Delta,q}(\mathbf{B})$ and that of complexity, $\Gamma_q(\mathbf{B})$ (see inset), at fixed length scales. On account of the high sensitivity of $\Gamma(k)$ for $k = 2, 3,$ and 4 (see Fig. 1a and b) these length scales were chosen for undertaking our comparison study. We do the same for the R pattern in Fig. 4b. Additionally, we consider other scales. The dashed curve corresponds to the smallest (first) length scale $k'$ at which the related measure vanishes for $q \to 0^+$. Here we find $k'(\mathrm{DSC}) = 10$ and $k'(\mathrm{R}) = 5$. Such behaviour indicates a lack of "empty" (with $n_i = 0$) cells of size equal or greater than $k' \times k'$ and may suggest some connection with geometrical parameters like "lacunarity", which measures the deviation of a fractal's behaviour from that of a translationally invariant one [21]. For a DSC pattern (see the inset in Fig. 1a), it suffices to note that it corresponds to the third stage of growth of the standard deterministic Sierpinski carpet: the central hole is smaller than at earlier stages of the growth and the actual pattern deviates from that of a translationally invariant lattice much less than its earlier forms (its lacunarity is, relatively, smaller). After applying one of the approximate expressions for lacunarity [22] to our two patterns the following effect can be observed: *the larger the pattern's lacunarity, the larger becomes the value of $k'$*. Moreover, for this length scale, and for larger ones as well, the duality $q \leftrightarrow (q_0)^2/q$, where $q_0$ denotes the coordinate of the peak of the entropic $q$-measure, is clearly detected [23]. We also remark on the fact that the two measures exhibit a peculiar behaviour at $k = 2$. Similar comments are to be made with reference to Fig. 5a and b, for $S_{\Delta,q}(\mathrm{PO}, \mathbf{B})$ and $\Gamma_q(\mathrm{PO}, \mathbf{B})$ measures. Note that the corresponding Shannon's values ($q = 1$) of the measures are located at the crossing between our curves and the vertical short-dashed line.

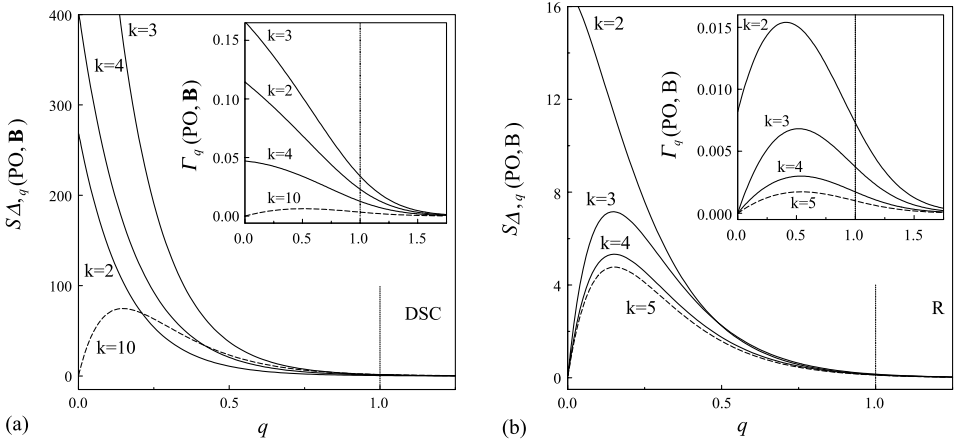

Fig. 5. The same as in Fig. 4 but for PO-measures.

The $q$-dependence investigated here allows for an additional interesting observation, related to FSO- and PO-measures of spatial inhomogeneity, which is illustrated in Fig. 6a. *At each scale, the corresponding FSO (solid) and PO (dotted-dashed) curves coincide for a finite and non-zero value of $q$, say, $q^*(k)$, that depends on the kind of pattern*. This means that for such a value of $q$, any geometrical constraint connected with the FSOs is somehow "translated" into the PO description. Thus, *each FSO can be treated as a PO*, which quite agrees with an intuitive picture of non-extensivity. Note that such a feature is absent in the case of the complexity measures investigated in this paper (Fig. 6b).

167

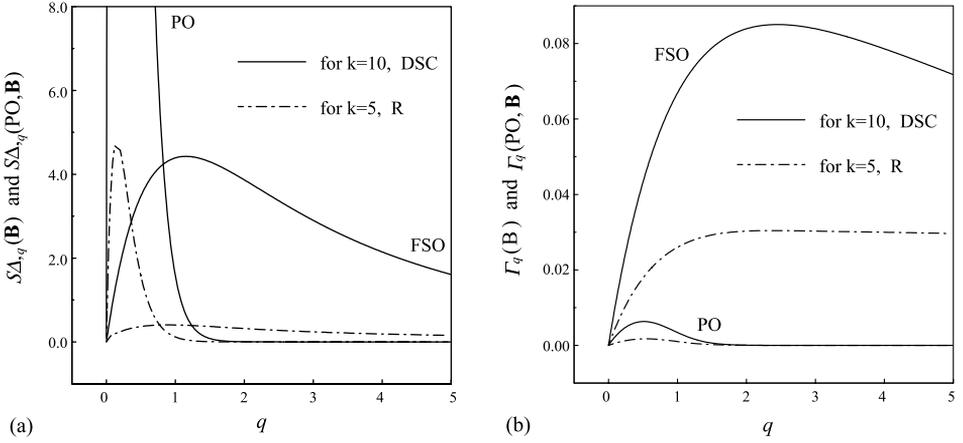

Fig. 6. Comparison of the inhomogeneity $q$-measures $S_{\Delta,q}(\mathbf{B})$ and $S_{\Delta,q}(\text{PO},\mathbf{B})$, at fixed length scales (a). The same for the complexity $q$-measures $\Gamma_q(\mathbf{B})$ and $\Gamma_q(\text{PO},\mathbf{B})$ (b). Note that the two concomitant FSO and PO inhomogeneity curves coincide for a specific, finite and nonzero value of $q$. This is not the case for the corresponding complexity curves.

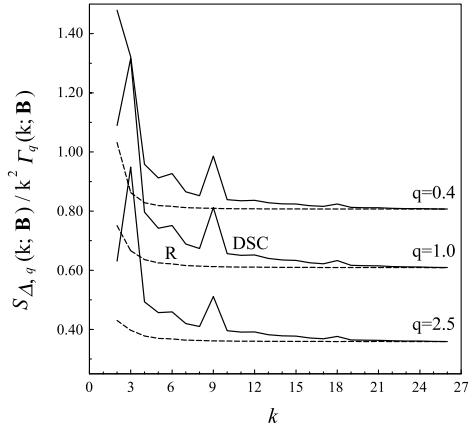

Fig. 7. The pairs (for DSC and R patterns, solid and dashed lines, respectively) of $S_{\Delta,q}(k;\mathbf{B})/k^2\Gamma_q(k;\mathbf{B})$ ratios for the Shannon case, $q=1$, and the Tsallis cases with: (a) fixed $q>1$ and (b) their counterparts $q<1$. We remark on the expected, unifying (converging) behaviour of both the ratios for large $k$-values.

In Fig. 7, we consider modified entropy/complexity ratios for DSC (solid lines) and R patterns (dashed curves). These $S_{\Delta,q}(k;\mathbf{B})/k^2\Gamma_q(k;\mathbf{B})$ ratios are evaluated for the Shannon case ($q=1$) and for the Tsallis instance as well. In the latter case we use, as in Fig. 3a, a fixed value $q=2.5$ and its counterpart $q=0.40$. We remark on the expected, unifying (converging) behaviour of these ratios for large $k$-values.

## 3. Conclusions

Our first remark concerns the nature of the approaches **A** and **B**. The **B**-technique seems to be more adequate for our purposes (that are mainly concerned with describing binary pixel patterns with FSOs) *because it is cell independent*. For this reason, method **B** was the one employed to compare (i) length scale's differences and (ii) $q$-differences between inhomogeneity and complexity measures.

For point (finite size) objects we detect a qualitative (quantitative indeed, for larger $k$-values) similarity, at different length scales, between $1/q$ and $q$-results (see Fig. 3a and b). (The $q \leftrightarrow (q_0)^2/q$ duality has been in fact specified and elucidated in Ref. [23].) With regard to the $q$-differences between inhomogeneity and complexity measures, let us focus out attention upon the most interesting interval, namely, $0 < q < 2$ and length scales $k \leqslant 10$. The maxima (at appropriate scales) of the FSO (PO) inhomogeneity measures concentrate around $q = 1$ (Fig. 4a, b, and 6a) and $q = 0.15$ (see Fig. 5a, b, and 6a) while, in the case of FSO complexity measures they are widely spread and rather smeared out (see, for instance, Fig. 6b or the insets in Fig. 4a and b). However, for the PO complexity measure the maxima are well defined (around $q = 0.50$, see the insets in Fig. 5a and b). On the other hand, the comparison of inhomogeneity and complexity measures illustrated in Fig. 4 and 5 still exhibits correlations that depend (weakly) on $q$.

The cell-dependence of configurational measures detected with the method **A** (see Eqs. (6,7) and Fig. 2a and b) strongly suggests that, for the specific system (binary pattern) considered here, the only valid microcanonical expression is Boltzmann's logarithmic one. Using instead the so-called $\ln_q$ measure does not reproduce the correct microcanonical results.

The present study should encourage further work along the lines tackled here. For instance, Abe and co-workers [24] have investigated a two-parameter $(q, q')$ family of statistical measures of complexity. These measures are based on a Tsallis-like, generalised Kulback–Leibner entropy [25–27]. In the limits $q, q' \to 1$ this Abe measure is equivalent (up to a constant prefactor) to that advanced in Refs. [12,13]. From a physical viewpoint it should be of interest to consider the generalised Kulback–Leibner entropy from the viewpoint of our present approach in order to make a more general comparison of the inhomogeneity and complexity $q$-measures. Additionally, a still open question is related to the physical meaning of the entropic index $q$ for binary patterns.

Recently [28–30], a meaningful relation between $q$ and the relative variance of locally fluctuating variables has been encountered, provided the fluctuating parameter is $\chi^2$ distributed. Given a binary pattern, the occupation numbers are distributed around the average number per cell. The deviation from the average depends on the structure. For instance, the relative variance is higher for a DSC than for the corresponding R pattern. Thus, if the occupation numbers are properly distributed, we can formally associate a $q$-parameter to such a pattern. This approach (although quite formal) may shed some light on the scale-dependent structural properties of binary patterns.

It is worth mentioning that, in a recent study [31,32], instead of using the difference $S_{\max} - S$ per cell, as we do here, attention is focused just on the entropy $S$ itself, with a different "$q$-modification" of the Shannon's and Tsallis forms. For $q < 1$, a connection between the lattice constant $\varepsilon$ (the phase space is partitioned into boxes of equal size) and the degree of non-extensitivity $1-q$ is discussed, under the assumption of equiprobability. In our language, for a strictly "*pattern-homogeneous*" distribution of POs (not phase space homogeneity), i.e., for $n_i = n/\chi$ for each $i$, the probabilities ($p_i = n_i/n$) defined in Appendix A become equal to $p_i = 1/\chi \equiv (k/L')^2$, which corresponds to $\varepsilon$ in Ref. [32]. Thus, in this specific case we find the relation $1 - q \sim (L')^2$, corresponding in our case to a two-dimensional pattern.

## Appendix A

In dealing with the PO-measure [10] one proceeds as follows:
(i) regards black pixels as punctual objects ($n_i$ per cell), while (ii) white pixels are treated as a continuous medium.

The appropriate Shannon-like forms are given by

$$S(k; \text{PO}) = \ln \left\{ \frac{n!}{n_1! n_2! \ldots n_\chi!} \right\} \cong -n \sum_{i=1}^{\chi} (n_i/n) \ln(n_i/n). \quad (A.1)$$

$$S_{\max}(k; \text{PO}) = \ln \left\{ \frac{n!}{(\chi - r_0) n_0! r_0(n_0 + 1)!} \right\}$$
$$\cong -n \left[ (\chi - r_0)(n_0/n) \ln(n_0/n) + r_0((n_0 + 1)/n) \ln((n_0 + 1)/n) \right] \quad (A.2)$$

and

$$S_\Delta(k; \text{PO}) = \frac{S_{\max}(k; \text{PO}) - S(k; \text{PO})}{\chi} \quad (A.3)$$

with $n_0$ and $r_0$ defined above. The associated Tsallis-like forms referred to as "the (**B**) approach" can be obtained by using again the Jackson derivative for each local cell

$$-D_q(n_i/n)^x|_{x=1} = [(n_i/n) - (n_i/n)^q]/(q-1). \quad (A.4)$$

yielding

$$S_q(k; \text{PO}, \mathbf{B}) = n \left[ 1 - \sum_{i=1}^{\chi} (n_i/n)^q \right] \Big/ (q-1) \quad (A.5)$$

and

$$S_{q,\max}(k; \text{PO}, \mathbf{B}) = n \left[ 1 - (\chi - r_0)(n_0/n)^q - r_0((n_0 + 1)/n)^q \right]/(q-1). \quad (A.6)$$

The (not used here) Tsallis-like forms for POs corresponding to the (**A**) approach can be obtained following the steps leading to Eqs. (6), (7), and using the proper number of possible microstates described in (A.1) and (A.2).

**Appendix B**

For the FSO-measure the rate (see below) $(*) \equiv (S_\Delta/k^2)/\Gamma = (S_{max})^2/(l^2 L^2 S)$ can be roughly estimated if we assume that

(i) $k$ is fixed and $L \bmod k = 0$, (thus $\chi \equiv L^2/k^2$ with $l = 1$, no periodic repetition of the initial pattern),
(ii) our configurations are "quasi-homogeneous" (i.e., for every $i$th cell we have $n_i \cong n_0$, with the total number of particles $n \cong \chi n_0$, and $r_0 \cong 0$, so that $S \cong S_{max}$),
(iii) the average number of black particles per cell, $n_0$, is big enough that we can use Stirling's approximation,
(iv) the concentration of black particles obeys the inequalities $0.5 < n/L^2 < 1$ (i.e., $x \equiv n_0/k^2 < 1$, $k^2/n_0 = 1 + (k^2 - n_0)/n_0 \equiv 1 + y$, with 1) $y < (1/x) - 1 < 1$, $\ln(1-x) \cong -x$, and 2) $\ln(1+y) \cong y$.

One easily obtains for the rate $(*)$:
$(*) \cong (S_{max})^2/(L^2 S_{max}) = S_{max}/L^2 = [\ln(k^2)! - \ln(n_0)! - \ln(k^2 - n_0)!]/k^2 \cong [k^2 \ln(k^2) - k^2 - n_0 \ln(n_0) + n_0 - (k^2 - n_0) \ln(k^2 - n_0) + k^2 - n_0]/k^2 = [n_0 \ln(1+y) - (k^2 - n_0) \ln(1-x)]/k^2 \cong 1$.